\documentclass[conference]{IEEEtran}
\usepackage{makecell}
\IEEEoverridecommandlockouts
\usepackage[linesnumbered,ruled,vlined]{algorithm2e}
\usepackage{booktabs}
\usepackage{multirow}
\usepackage{cite}
\usepackage{amsmath,amssymb,amsfonts}
\usepackage{algorithmic}
\usepackage{graphicx}
\usepackage{textcomp}
\usepackage{xcolor}
\usepackage{comment}
\usepackage{hyperref}
\usepackage{soul}
\usepackage{placeins}
\usepackage{dblfloatfix}
\usepackage{cuted}
\usepackage{capt-of}

\usepackage{float}

\usepackage{booktabs} % For nice horizontal rules
\usepackage{array}    % For extended column definitions (optional)
\begin{document}
\title{AutoPoly-HE: Autonomous Design of Polynomial Activations for Privacy-Preserving Encrypted Inference}

\author{
\IEEEauthorblockN{
Zeinab Elkhatib$^{1}$, Kamrul Hasan$^{1}$, Ali Sekmen$^{1}$, Imtiaz Ahmed$^{2}$, Tariqul Islam$^{3}$
}

\IEEEauthorblockA{
$^{1}$ Tennessee State University, Nashville, TN, USA \\
%$^{2}$ Old Dominion University, Norfolk, VA, USA \\
$^{2}$ Howard University, Washington, DC, USA \\
$^{3}$ University of Maryland Baltimore County (UMBC), Baltimore, MD, USA \\
Email: $\lbrace$\textit{zelkhati, mhasan1, asekmen }$\rbrace$@tnstate.edu, \\
%\textit{sshetty}@odu.edu, 
\textit{imtiaz.ahmed}@howard.edu, 
\textit{mtislam}@umbc.edu
}
}
\maketitle

\begin{abstract}

 Privacy-preserving machine learning has become critical for deploying artificial intelligence in sensitive domains such as healthcare, finance, and infrastructure monitoring. Homomorphic encryption (HE) enables secure inference on encrypted data. Despite its advantages, HE is not compatible with nonlinear activation functions found in machine learning models. Existing methods mitigate this problem by manually designing polynomial approximations for these activation functions. However, these functions do not account for the differences in dataset characteristics or model architecture. Furthermore, these functions are also limited to the accuracies of convolutional neural network (CNN) models. In this work, we propose an autonomous framework that selects polynomial activation functions tailored to each dataset and model. Using this framework, we analyze the interdependence between polynomial activation design and model architecture by evaluating several CNN ResNet and Vision Transformer models across multiple datasets. Experimental results demonstrate that the proposed framework generalizes across multiple architectures, including ResNet-18, ResNet-34, ResNet-50, ViT-S/16, ViT-B/16, and ViT-L/16. Across all datasets, transformer-based models maintain higher accuracy under polynomial approximation, while convolutional models exhibit increased degradation as task complexity grows. In addition, latency analysis reveals that the proposed CLS token based inference pipeline significantly reduces computational overhead by eliminating depth-dependent operations, shifting the primary cost to encryption and decryption. These findings demonstrate that the proposed method not only adapts polynomial activations across architectures, but also enables efficient encrypted inference through both activation design and pipeline optimization.

\end{abstract}

\section{Introduction}

The increasing demand for secure artificial intelligence has driven interest in privacy-preserving machine learning (PPML), particularly in domains such as healthcare, autonomous transportation, and other critical infrastructures, where sensitive data must remain protected \cite{one}, \cite{two}. In many cases, machine learning inference is performed on external or cloud-based systems, creating a need for techniques that allow computation without exposing raw data \cite{three}. 

Homomorphic Encryption (HE) provides a solution by enabling computation directly on encrypted data \cite{four}. However, the application of HE in deep learning is limited by the types of operations it supports. Homomorphic encryption schemes only support linear operations, such as addition and multiplication. This makes it infeasible to directly evaluate nonlinear activation functions, including ReLU and GELU, which are essential for modern neural networks \cite{five}, \cite{six}. To address these limitations, these functions are replaced with polynomial approximations that are compatible with HE. While this enables encrypted computation, it introduces approximation errors that can significantly degrade model accuracy. 

Furthermore, although polynomial activations make it possible to use encryption with machine learning, achieving high accuracy remains a major challenge. Particularly with more complex datasets, models often experience performance degradation. This suggests that the impact of polynomial approximations extends beyond function design alone and depends on additional factors within the model. Two key challenges contribute to this limitation. First, polynomial activation functions are typically designed by approximating standard nonlinear functions such as ReLU or GELU directly, rather than being tailored to the characteristics of the dataset or the behavior of the trained model. This approach can result in suboptimal approximations that do not generalize well throughout the model. The second challenge limiting the accuracy of encrypted inference is the role of model architecture, specifically how it interacts with polynomial approximation. Most efforts have focused on CNNs, without investigating how alternative architectures, such as ViTs, behave under these constraints. To address these challenges, we introduce an automated framework that learns polynomial activation functions based on the dataset and trained model, and that automates the search for the appropriate polynomial degree. This removes the need for manual design and helps maintain a balance between accuracy and computational efficiency.

Using this framework, we analyze the interaction between polynomial activation design and model architecture by evaluating multiple convolutional and transformer-based models, including ResNet-18, ResNet-34, ResNet-50, ViT-S/16, ViT-B/16, and ViT-L/16 across multiple datasets. To address these challenges, the proposed framework introduces three key improvements: a) it learns polynomial activation functions tailored to the dataset and trained model, reducing approximation error; b) it balances accuracy and homomorphic encryption constraints by selecting polynomial degrees that control multiplicative depth; and c) it evaluates both convolutional and transformer-based architectures, enabling analysis of how model structure impacts performance under polynomial approximations.
The rest of this paper is organized as follows. Section II reviews existing work on privacy-preserving machine learning, homomorphic encryption, and polynomial activation functions. Section III describes the proposed methodology, including the automated framework for generating polynomial activations and the experimental setup used for both CNN and the Vision Transformer model. Section IV presents and analyzes the experimental results, highlighting the impact of activation design and model architecture on performance. Finally, Section V concludes the paper and outlines directions for future work.

\section{Related Work}

Prior work on privacy-preserving semantic communication highlights four related challenges: selecting task-relevant image information for transmission, letting the server use that information without seeing its unencrypted values, adapting models for computation on encrypted data, and evaluating these choices across datasets and architectures.
The first challenge concerns what to transmit. Semantic communication sends information needed for a task instead of the complete original input. DeepSC demonstrated this principle for text transmission \cite{deepsc}. For visual data, Yang et al. applied semantic communication to image classification and surface-defect detection \cite{yang_semantic}, while Lyu et al. used transmitted image features for both classification and image recovery \cite{lyu_semantic}. These studies motivate sending task-relevant image representations. However, the representation itself can reveal private information. Wang et al. studied model inversion attacks that attempt to reconstruct original data from transmitted task-related features \cite{wang_privacy_semantic}. Reducing the amount of image data sent therefore does not, by itself, prevent the server from learning about the image.
The second challenge is letting the server classify the transmitted features without seeing their unencrypted values. Homomorphic encryption (HE) allows a server to compute on encrypted data and return an encrypted result for the client to decrypt. Prior HE-based inference systems demonstrate the feasibility of this approach \cite{cryptonets,gazelle,highly,redsec}. However, many evaluations use benchmark datasets such as MNIST and CIFAR-10. Their results do not fully establish how an encrypted client-to-server system performs with more varied image data, or how the size of an encrypted intermediate representation affects communication cost.
The third challenge is adapting the model for encrypted computation. Standard nonlinear activations cannot be evaluated directly using the arithmetic supported by common HE schemes, so HE inference studies often replace them with polynomial approximations \cite{highly,low,aespa}. Choosing a polynomial involves a trade-off: its degree and evaluation method affect the required multiplicative depth and computation cost, while the approximation can change classification accuracy. For example, \cite{highly} reports CIFAR-10 accuracies ranging from 76.67\% to 81.06\% across the approximations it evaluates. Automated methods seek to reduce the manual effort involved in choosing HE-compatible activations \cite{ola,slaf,autofhe}; Table~\ref{tab:related_work_small} compares these approaches.
The fourth challenge is understanding how model architecture affects encrypted semantic transmission. ResNets generate spatial feature maps through convolutional layers \cite{he2016resnet}, whereas Vision Transformers convert image patches into tokens and process them as a sequence \cite{han2023vit}. If inference is divided between a client and a server, the client must transmit the intermediate representation produced at the split point. Prior work shows that selecting this point changes the computation assigned to the client and the amount of data sent to the server \cite{bottlenetpp}. Under HE, the size of that representation also matters because it determines how much data must be encrypted and transmitted, while the operations in the remaining layers determine the difficulty of encrypted inference. We evaluate ResNet and Vision Transformer models on CIFAR-10, CIFAR-100, and medical images to examine how architecture and dataset affect classification accuracy and encrypted inference cost.
\begin{table}[t]
\centering
\caption{Summary of prior HE-based inference methods.}
\label{tab:related_work_small}
\scriptsize
\setlength{\tabcolsep}{3pt}

\begin{tabular}{p{2.6cm} p{2.0cm} p{2.0cm} p{1.2cm}}
\toprule
\textbf{Paper} & \textbf{Dataset} & \textbf{Activation} & \textbf{Acc.} \\
\midrule

CryptoNets
\cite{cryptonets} & MNIST & Manual & 90\% \\

Redsec \cite{redsec} & MNIST, CIFAR-10 & Manual & 85\% \\

Highly \cite{highly} & CIFAR-10 & Manual & 81.06\% \\

OLA \cite{ola} & CIFAR-10 & Automatic & 92.5\% \\

SLAF \cite{slaf} & MNIST & Automatic & 99.38\% \\

AutoFHE \cite{autofhe} & CIFAR-10 & Automatic & 92.68\% \\

\bottomrule
\end{tabular}
\end{table}

This last point is the third gap: Vision Transformers now outperform CNNs on complex visual tasks \cite{imagenet}, \cite{vit}, \cite{deit}, \cite{swin}, yet their integration with HE remains largely unexplored -- existing privacy-preserving ViT work instead relies on secure multi-party computation \cite{mpcvit}, \cite{privit}, \cite{shaft} rather than full HE, due to the added computational depth transformers introduce.

Beyond privacy-preserving computation, a parallel line of work asks what should be transmitted in the first place. Semantic communication reframes transmission around task-relevant meaning rather than raw bits, using learned encoders to compress an input into a compact representation optimized for a downstream task rather than bit-level fidelity \cite{semcom}, \cite{semcom_survey}. Such representations, though far smaller than the raw input, can still leak sensitive information if intercepted -- a security question largely unaddressed in that literature. This work sits at the intersection of these gaps: an autonomous framework that learns HE-compatible polynomial activations tailored to dataset and model, evaluated across both CNN and ViT architectures, enabling the compact semantic representations these models produce to be encrypted and transmitted without hand-tuned, architecture-specific approximations.

\section{Methodology}

\begin{figure*}[h!]
    \centering
    \includegraphics[width=0.8\textwidth,height=0.40\textheight]{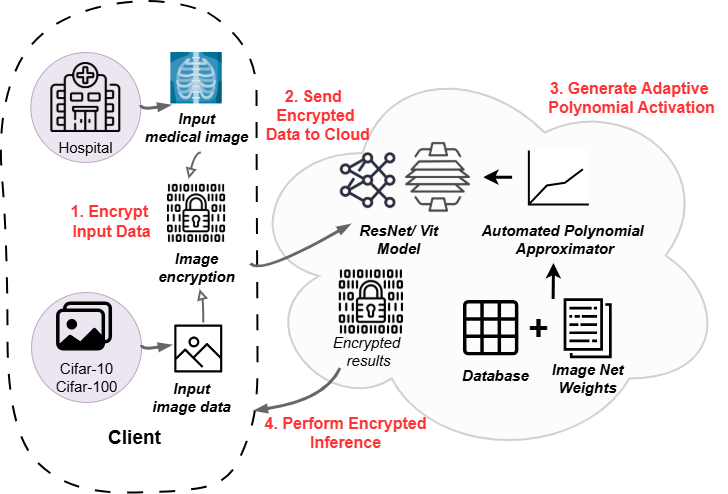}
    \caption{Complete Model Described}
    \label{fig:relu}
\end{figure*}

To simplify the construction of homomorphic encryption compatible activation functions, we developed an automated framework, which is designed to work across different model types. In this work, the framework is applied across multiple architectures, including ResNet-18, ResNet-34, ResNet-50, ViT-S/16, ViT-B/16, and ViT-L/16, to evaluate scalability and generalization. All experiments were conducted on a system equipped with an Intel\textsuperscript{\textregistered} Core\textsuperscript{TM} i9-14900HX CPU (24 cores, 32 threads, base clock 2.20~GHz) and 31.71~GB of RAM. Performance is assessed in terms of classification accuracy and inference time. An overview of the proposed framework is illustrated in the Figure \ref{fig:relu}.

\subsection{The Automated Polynomial Activation Framework}

In this work, we propose an automated framework for constructing homomorphic encryption compatible activation functions, where all nonlinear activations are replaced with functions that can be evaluated using only addition and multiplication. The framework is designed to learn activation functions directly from data and pretrained models, eliminating the need for manual approximation. We first start by defining the neural network as \cite{nn}:

\begin{equation}
\hat{y} = f_{\theta}(x)
\end{equation}
where $x$ is the input, $\theta$ represents the model parameters, and $\hat{y}$ is the predicted output. The goal is to replace each nonlinear activation function $\sigma(x)$ with a polynomial function $P(x)$, such that:
\begin{equation}
\sigma(x) \rightarrow P(x)
\end{equation}

The polynomial activation function is defined as:
\begin{equation}
P(x) = \sum_{i=0}^{d} a_i x^i
\end{equation}
where $d$ is the polynomial degree and $a_i$ are learnable coefficients. To ensure stable initialization, the coefficients are initialized as:
\begin{equation}
a_1 = 1, \quad a_i = 0 \text{ for } i \neq 1
\end{equation}
allowing the activation function to initially behave as an identity mapping. Next, to improve approximation accuracy and numerical stability, clipping is introduced to constrain the input range:
\begin{equation}
x_{\text{clip}} = \text{clamp}(x, -T, T)
\end{equation}
The clipping threshold $T$ is defined as \cite{softplus7}:
\begin{equation}
T = \text{softplus}(t_{\text{raw}}) + \epsilon
\end{equation}
where:
\begin{equation}
\text{softplus}(u) = \ln(1 + e^u)
\end{equation}
and $\epsilon$ is a small positive constant. This ensures that $T > 0$ while allowing it to be learned during training. The activation function is therefore applied as:
\begin{equation}
P(x_{\text{clip}}) = \sum_{i=0}^{d} a_i (x_{\text{clip}})^i
\end{equation}

To reduce computational complexity and multiplicative depth, the polynomial is evaluated using Horner’s method \cite{dorn1962generalizations}:
\begin{equation}
P(x) = a_0 + x(a_1 + x(a_2 + \cdots + x a_d))
\end{equation}

The framework is then trained using labeled datasets to learn both the network parameters and the activation functions simultaneously. Where the training dataset is defined as:
\begin{equation}
D = \{(x_i, y_i)\}_{i=1}^{N}
\end{equation}

In this work, three datasets are used to evaluate the framework: CIFAR-10\cite{resnet18_cifar100}, CIFAR-100\cite{resnet18_cifar100}, and a brain tumor MRI dataset\cite{nawaz2022brain}. These datasets are selected to represent varying levels of complexity and class diversity, allowing the framework to be evaluated across different data distributions.

The model is then trained using the cross-entropy loss:
\begin{equation}
\mathcal{L}_{CE} = - \sum_{i=1}^{N} \sum_{c=1}^{C} y_{i,c} \log(\hat{y}_{i,c})
\end{equation}

During training, both the network parameters $\theta$ and the polynomial coefficients $\{a_i\}$ are optimized jointly using gradient-based optimization:
\begin{equation}
\theta_{t+1} = \theta_t - \eta \nabla_{\theta} \mathcal{L}_{CE}
\end{equation}

And a cosine annealing learning rate schedule \cite{cosine} is applied to improve convergence:
\begin{equation}
\eta_t = \eta_{\min} + \frac{\eta_0 - \eta_{\min}}{2} \left(1 + \cos\left(\frac{\pi t}{T_{\max}}\right)\right)
\end{equation}

The key component of this framework is the automatic selection of the polynomial degree. Instead of fixing the degree manually, multiple candidate degrees are evaluated:
\begin{equation}
d \in \{2, 3, 4, 5, 6\}
\end{equation}

For each degree $d$, the model is trained on the dataset and its classification accuracy is computed as:
\begin{equation}
\text{Acc}(d) = \frac{\text{correct predictions}}{\text{total samples}}
\end{equation}

Then, to account for homomorphic encryption constraints, the multiplicative depth required to evaluate the polynomial is approximated as:
\begin{equation}
\text{Depth}(d) = \lceil \log_2(d) \rceil
\end{equation}

Next, a scoring function is defined to balance model accuracy and homomorphic evaluation cost. For CNN-based architectures, both network depth and polynomial degree contribute to computational complexity due to repeated nonlinear evaluations across layers. This is captured by:

\[
\text{Score}(d) = \text{Acc}(d) - \alpha \cdot \text{Depth} - \beta \cdot (d - 1)
\]

where $\alpha = 0.005$ and $\beta = 0.0005$ control the trade-off between accuracy and computational cost.

In contrast, for the proposed CLS-token-based transformer inference pipeline, deep sequential operations are eliminated, significantly reducing the impact of network depth. As a result, the scoring function simplifies to:

\[
\text{Score}(d) = \text{Acc}(d) - \beta \cdot (d - 1)
\]

reflecting that polynomial degree remains the primary contributor to homomorphic computation cost in this setting.

Then, the optimal polynomial degree is selected as:
\begin{equation}
d^* = \arg\max_d \text{Score}(d)
\end{equation}

Once the optimal degree is determined, the corresponding polynomial coefficients and clipping thresholds are extracted. For each activation layer $l$, the learned activation function is:
\begin{equation}
P_l(x) = \sum_{i=0}^{d^*} a_i^{(l)} \left( \text{clamp}(x, -T_l, T_l) \right)^i
\end{equation}

Finally, The resulting activation function consists only of linear operations, while balancing computational power and accuracy, making it fully compatible with homomorphic encryption. Once the optimal approximation is determined, it is integrated into both the ResNet and Vision Transformer models, as described in the following sections.

\subsection{The Three Datasets}

In this work, three datasets are used to evaluate the proposed framework: CIFAR-10, CIFAR-100, and a brain tumor MRI dataset. These datasets represent varying levels of task complexity and real-world relevance. CIFAR-10 consists of 60,000 color images, $32 \times 32$, across 10 classes, with 50,000 training and 10,000 testing samples. Due to its limited number of classes, it serves as a baseline for simpler classification tasks. CIFAR-100 follows the same structure but includes 100 classes with fewer samples per class. The brain tumor MRI dataset contains medical images categorized into four classes: glioma, meningioma, pituitary tumor, and no tumor. This dataset represents a real-world, privacy-sensitive application with greater variability and complexity. These datasets enable evaluation across low-complexity, higher-complexity, and real-world (MRI) scenarios, providing a comprehensive assessment of the proposed framework.

\subsection{Application to CNN Architectures: ResNet}

The proposed framework is applied to ResNet-18, ResNet-34, and ResNet-50 across CIFAR-10, CIFAR-100, and a brain tumor MRI dataset. The final classification layer is adjusted to match the number of classes for each dataset. All ReLU activations are replaced with learned polynomial functions obtained from the proposed framework. Models are initialized with pretrained ImageNet weights and trained under standard supervised settings, allowing the impact of polynomial activations to be isolated. Furthermore, models are also trained and evaluated using polynomial degrees ranging from 2 to 6 to analyze the impact of polynomial approximation on CNN performance. For encrypted inference, the framework selects the optimal polynomial activation from degrees 2 to 6 based on a balance between accuracy and computational cost, and evaluates classification under homomorphic encryption using only addition and multiplication.

\subsection{Application to Transformer Architectures: Vision Transformers}

To evaluate the proposed hybrid framework on transformer-based models, it is applied to ViT-S/16, ViT-B/16, and ViT-L/16, across the same three datasets. The classification head is adjusted to match the number of output classes, and all GELU activations are replaced with learned polynomial functions obtained from the proposed framework. Models are initialized with pretrained ImageNet weights, from the client, and trained under standard supervised settings. For encrypted inference, a CLS-token-based pipeline is adopted to reduce the computational cost. In this approach, the transformer model processes inputs in plaintext to extract the CLS token representation, which summarizes the image features. Only this compact representation is then encrypted and used for the final classification step, avoiding expensive layer-by-layer computation under homomorphic encryption. The selected polynomial activation, is then applied using only addition and multiplication, ensuring compatibility with homomorphic encryption.

\subsection{Encrypted Inference Parameters}

Encrypted inference is performed using the CKKS scheme implemented in TenSEAL \cite{tenseal}, which supports approximate arithmetic on real-valued data. Nonlinear activation functions are replaced with polynomial approximations, where multiplicative depth is determined by the polynomial degree. A standard configuration ($N = 8192$, coefficient modulus $[60, 40, 40, 60]$, scale $2^{40}$) supports evaluation up to degree 4. For more complex settings, a larger configuration ($N = 16384$, $[60, 40, 40, 40, 60]$, scale $2^{45}$) improves numerical stability. This design enables efficient and stable encrypted inference across all evaluated architectures.

\section{Results \& Discussions}
\vspace{-0.1em}

The results demonstrate that the proposed framework generalizes across both convolutional and transformer-based architectures, while revealing clear differences in how models respond to polynomial approximation. Across all datasets, As shown in Table~\ref{tab:combined_perf_latency_full}, transformer-based models, ViT-S/16, ViT-B/16, and ViT-L/16, consistently achieve higher test accuracy than their ResNet counterparts. On CIFAR-10 and the MRI dataset, both Resnet and transformer models maintain high accuracy across polynomial degrees. However,  with CIFAR-100, CNN-based models exhxibited more noticeable performance degradation, this performance gap becomes more pronounced, with deeper CNNs (ResNet-34 and Resnet-50) failing to match the accuracy achieved by transformer-based models. 

\begin{strip}
\centering
\tiny
\captionof{table}{Performance, selected polynomial activations, and encrypted inference latency across datasets and architectures.}
\label{tab:combined_perf_latency_full}
\setlength{\tabcolsep}{2pt}
\renewcommand{\arraystretch}{0.95}

\resizebox{\textwidth}{!}{
\begin{tabular}{llccc l cc cc cc cc c}
\toprule

% ===== TOP GROUP HEADERS =====
\multicolumn{6}{c}{Performance and Polynomial Activation of Models} 
& \multicolumn{8}{c}{Encrypted Inference Latency Calculation} \\

\cmidrule(lr){1-6} \cmidrule(lr){7-14}

% ===== SECOND HEADER =====
& & & & & & \multicolumn{2}{c}{Encryption} 
& \multicolumn{2}{c}{Decryption} 
& \multicolumn{2}{c}{Prototype / FC} 
& \multicolumn{2}{c}{Polynomial} & \\

\cmidrule(lr){7-8} \cmidrule(lr){9-10} \cmidrule(lr){11-12} \cmidrule(lr){13-14}

Dataset & Model & Deg. & Train & Test & Polynomial 
& s & \% & s & \% & s & \% & s & \% & Total \\
\midrule

% ================= CIFAR-10 =================
CIFAR-10 & ResNet-18 & 2 & 86.48 & 85.93 &
$0.2339x^2+1.1185x-0.0533$
& 0.01 & 1.52 & 0.02 & 3.06 & 0.32 & 47.64 & 0.32 & 47.77 & 0.67 \\

CIFAR-10 & ResNet-34 & 2 & 88.56 & 87.43 &
$0.1927x^2+0.55x+0.2244$
& 0.012 & 1.3 & 0.025 & 2.7 & 0.52 & 48.0 & 0.53 & 48.0 & 1.09 \\

CIFAR-10 & ResNet-50 & 4 & 88.89 & 88.67 &
$-0.0252x^4+0.012x^3+0.2929x^2+0.55x+0.1458$
& 0.015 & 1.1 & 0.030 & 2.3 & 0.80 & 48.3 & 0.82 & 48.3 & 1.67 \\

CIFAR-10 & ViT-S/16 & 2 & 91.83 & 90.12 &
$0.1862x^2+0.425x+0.0386$
& 0.0060 & 76.5 & 0.0017 & 21.7 & 0.00007 & 0.9 & 0.00002 & 0.2 & 0.00779 \\

CIFAR-10 & ViT-B/16 & 3 & 94.94 & 94.77 &
$0.0421x^3+0.2194x^2+0.2700x-0.0322$
& 0.00815 & 77.5 & 0.00221 & 21.0 & 0.00010 & 1.0 & 0.00003 & 0.3 & 0.01052 \\

CIFAR-10 & ViT-L/16 & 2 & 94.98 & 94.02 &
$0.1453x^2+0.425x+0.0777$
& 0.0145 & 78.0 & 0.0040 & 21.5 & 0.00018 & 0.9 & 0.00006 & 0.3 & 0.01874 \\

\midrule

% ================= CIFAR-100 =================
CIFAR-100 & ResNet-18 & 3 & 58.58 & 57.48 &
$-0.0582x^3+0.2018x^2+1.1467x-0.1079$
& 0.009 & 0.13 & 0.19 & 2.62 & 3.01 & 40.45 & 4.22 & 56.8 & 7.44 \\

CIFAR-100 & ResNet-34 & 4 & 62.47 & 61.09 &
$-0.0078x^4+0.1944x^2+0.4750x+0.3771$
& 0.011 & 0.1 & 0.25 & 2.3 & 4.80 & 44.5 & 5.70 & 53.1 & 10.76 \\

CIFAR-100 & ResNet-50 & 4 & 67.53 & 67.01 &
$-0.0158x^4+0.0120x^3+0.1777x^2+0.4750x+0.3916$
& 0.013 & 0.1 & 0.35 & 2.1 & 7.20 & 43.5 & 9.00 & 54.3 & 16.56 \\

CIFAR-100 & ViT-S/16 & 4 & 83.09 & 83.01 &
$-0.0082x^4+0.217x^2+0.45x+0.1346$
& 0.0085 & 75.8 & 0.0023 & 20.5 & 0.00025 & 2.2 & 0.00004 & 0.3 & 0.01109 \\

CIFAR-100 & ViT-B/16 & 4 & 86.00 & 86.09 &
$-0.009484x^4+0.0084x^3+0.2511x^2+0.4669x+0.1337$
& 0.01219 & 76.4 & 0.00330 & 20.7 & 0.00038 & 2.4 & 0.00005 & 0.3 & 0.01595 \\

CIFAR-100 & ViT-L/16 & 4 & 86.98 & 86.15 &
$-0.0151x^4+0.0120x^3+0.1792x^2+0.45x+0.1532$
& 0.0215 & 77.5 & 0.0058 & 21.0 & 0.00070 & 2.5 & 0.00008 & 0.3 & 0.02808 \\

\midrule

% ================= MRI =================
MRI & ResNet-18 & 2 & 90.19 & 89.64 &
$0.0395x^2+1.0289x+0.0326$
& 0.009 & 3.26 & 0.009 & 3.18 & 0.11 & 41.64 & 0.14 & 51.85 & 0.28 \\

MRI & ResNet-34 & 2 & 90.96 & 89.87 &
$0.1362x^2+0.5x+0.1512$
& 0.011 & 2.8 & 0.011 & 2.8 & 0.18 & 45.0 & 0.20 & 49.4 & 0.40 \\

MRI & ResNet-50 & 4 & 92.56 & 91.28 &
$-0.0151x^4+0.012x^3+0.1852x^2+0.5x+0.1511$
& 0.012 & 2.5 & 0.012 & 2.5 & 0.28 & 46.0 & 0.31 & 49.0 & 0.61 \\

MRI & ViT-S/16 & 4 & 90.34 & 90.02 &
$-0.0055x^4+0.1949x^2+0.5400x+0.049$
& 0.0052 & 77.0 & 0.0014 & 20.8 & 0.00006 & 0.9 & 0.00003 & 0.4 & 0.00669 \\

MRI & ViT-B/16 & 2 & 93.93 & 91.94 &
$0.2918x^2+0.4984x+0.0390$
& 0.00748 & 77.9 & 0.00198 & 20.6 & 0.00009 & 0.9 & 0.00005 & 0.5 & 0.00960 \\

MRI & ViT-L/16 & 4 & 94.56 & 93.97 &
$-0.0137x^4+0.012x^3+0.166x^2+0.514x+0.51$
& 0.0130 & 78.2 & 0.0036 & 21.6 & 0.00016 & 1.0 & 0.00007 & 0.4 & 0.01683 \\

\bottomrule
\end{tabular}
}
\end{strip}

\begin{figure}[H]
    \centering
    \includegraphics[width=\columnwidth]{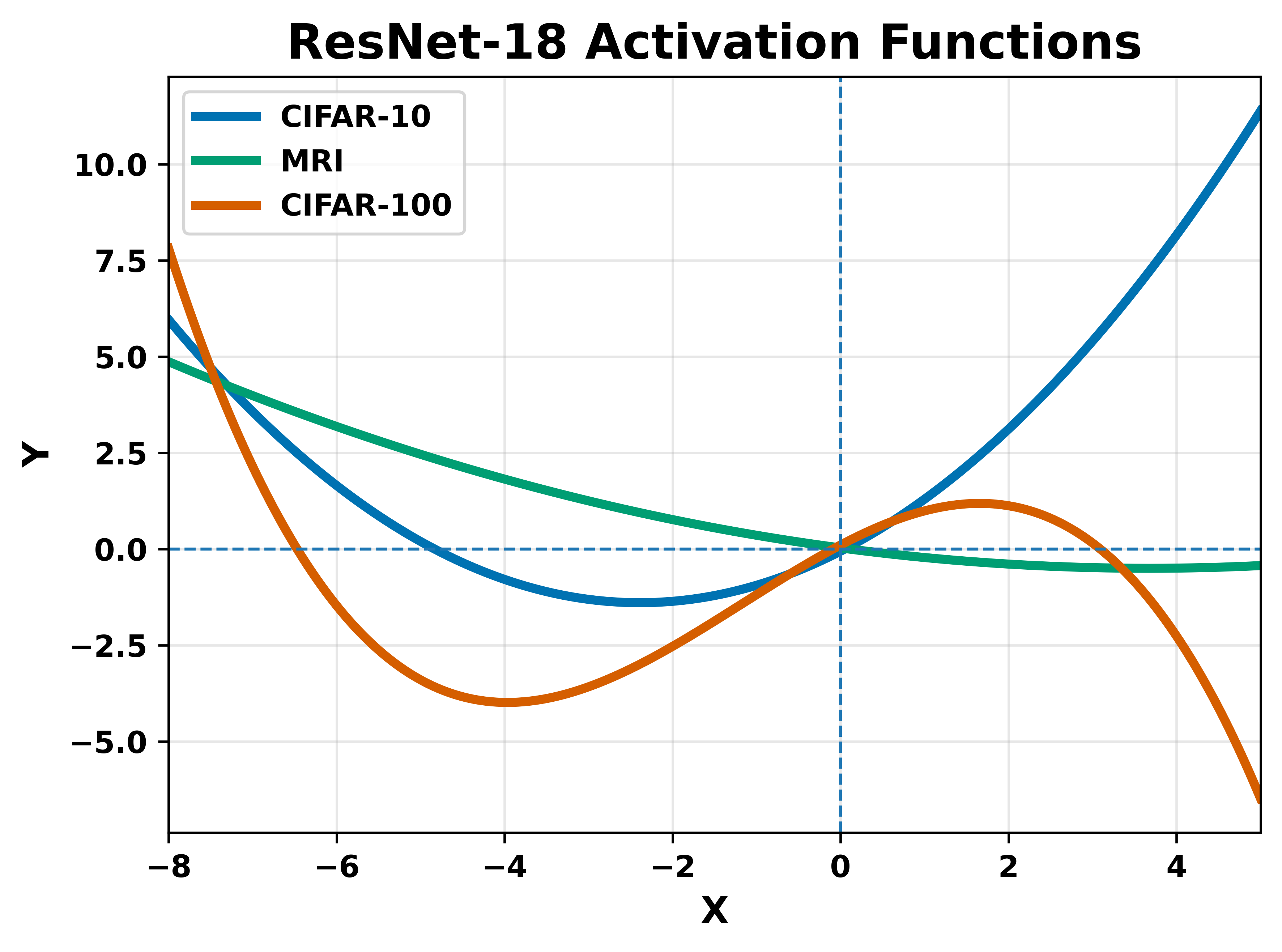}
    \caption{Learned polynomial activation functions for ResNet-18 across CIFAR-10, MRI, and CIFAR-100.}
    \label{fig:resnet18_activations}
\end{figure}

\begin{figure}[H]
    \centering
    \includegraphics[width=\columnwidth]{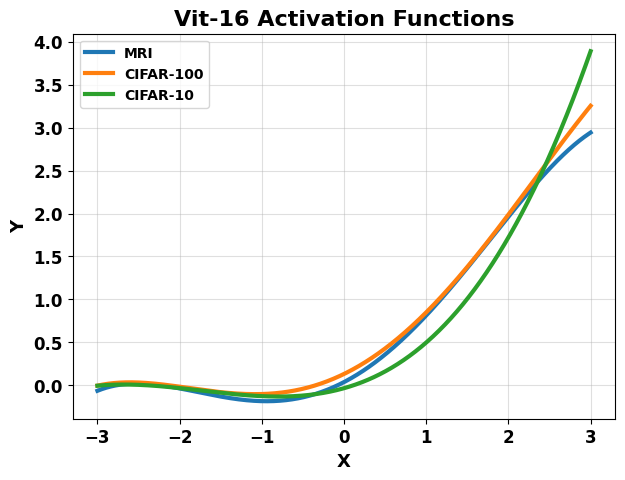}
    \caption{Learned polynomial activation functions for ViT-16 across CIFAR-10, MRI, and CIFAR-100.}
    \label{fig:vit16_activations}
\end{figure}

These trends are further illustrated in Fig.~\ref{fig:vit16_activations} and Fig.~\ref{fig:resnet18_activations}, where the learned polynomial activation functions show smoother and more stable behavior for transformer-based models, while CNN-based models exhibit greater distortion under polynomial constraints.
In addition to accuracy, Table~\ref{tab:combined_perf_latency_full}, latency results reveal a significant computational advantage for transformer-based models. CNN-based architectures include substantial overhead due to repeated polynomial evaluations across multiple layers, whereas the proposed CLS-token-based pipeline eliminates depth-dependent computation. As a result, transformer-based inference is dominated by encryption and decryption time, leading to significantly lower total latency compared to CNN-based approaches. These findings demonstrate that the proposed framework not only adapts polynomial activations across architectures, but also enables efficient encrypted inference through both activation selection and an optimized inference pipeline. This is likely due to the absence of architecture specific optimizations and the use of a generalized framework designed for multiple datasets, which prioritizes flexibility over peak performance. Finally, Although ViT-16 achieves higher accuracy, it requires greater computational cost than ResNet-18 due to the use of higher-degree polynomial activations, which increase the multiplicative depth in homomorphic encryption.

\subsection*{Future Discussion}
The results on CIFAR-100 motivate further exploration of scalability. Future work will extend the framework beyond vision tasks and leverage agent-based methods to automate additional components of model and inference design.

\section{Acknowledgement}

This work is supported by the National Science
Foundation (NSF) Award number 2409093.

%\bibliographystyle{IEEEtran}
%\bibliography{references}

\bibliographystyle{ieeetr}
\bibliography{references.bib}

@article{one,
  title={Privacy-Preserving Machine Learning: Threats and Solutions},
  author={Yang, Qiang and Liu, Yang and Chen, Tianjian and Tong, Yongxin},
  journal={ACM Computing Surveys},
  year={2021}
}

@article{two,
  title={A Review of Privacy-Preserving Techniques for Medical Data Sharing},
  author={Rieke, Nicola and Hancox, Jessica and Li, Wenqi and Milletari, Fausto and Roth, Holger and Albarqouni, Shadi and Bakas, Spyridon and others},
  journal={Nature Machine Intelligence},
  year={2020}
}

@article{three,
  title={Machine Learning in the Cloud: Security and Privacy Challenges},
  author={Zhang, Qian and Chen, Min and others},
  journal={IEEE Network},
  year={2019}
}

@inproceedings{four,
  title={Homomorphic Encryption for Arithmetic of Approximate Numbers},
  author={Cheon, Jung Hee and Kim, Andrey and Kim, Miran and Song, Yongsoo},
  booktitle={Advances in Cryptology – ASIACRYPT},
  year={2017}
}

@inproceedings{five,
  title={CryptoNets: Applying Neural Networks to Encrypted Data with High Throughput and Accuracy},
  author={Dowlin, Nathan and Gilad-Bachrach, Ran and others},
  booktitle={ICML},
  year={2016}
}

@inproceedings{six,
  title={CryptoDL: Deep Neural Networks over Encrypted Data},
  author={Hesamifard, Ehsan and Takabi, Hassan and Ghasemi, Mehdi},
  booktitle={arXiv preprint arXiv:1711.05189},
  year={2017}
}

@inproceedings{cryptonets,
  title={CryptoNets: Applying Neural Networks to Encrypted Data with High Throughput and Accuracy},
  author={Dowlin, Nathan and Gilad-Bachrach, Ran and Laine, Kim and Lauter, Kristin and Naehrig, Michael and Wernsing, John},
  booktitle={Proceedings of the 33rd International Conference on Machine Learning (ICML)},
  year={2016}
}

@inproceedings{gazelle,
  title={Gazelle: A Low Latency Framework for Secure Neural Network Inference},
  author={Juvekar, Chiraag and Vaikuntanathan, Vinod and Chandrakasan, Anantha},
  booktitle={USENIX Security Symposium},
  year={2018}
}

@INPROCEEDINGS{highly,
  author={Ishiyama, Takumi and Suzuki, Takuya and Yamana, Hayato},
  booktitle={2020 IEEE International Conference on Big Data (Big Data)}, 
  title={Highly Accurate CNN Inference Using Approximate Activation Functions over Homomorphic Encryption}, 
  year={2020},
  volume={},
  number={},
  pages={3989-3995},
  doi={10.1109/BigData50022.2020.9378372}}

@inproceedings{redsec,
  title={REDsec: Running Encrypted Deep Neural Networks in Secure Enclaves},
  author={Lee, Joohee and others},
  booktitle={Proceedings of the ACM Conference on Computer and Communications Security (CCS)},
  year={2021}
}

@inproceedings{low,
  title={LoLa: Low-Latency CryptoNets for Homomorphic Inference},
  author={Brutzkus, Alon and Elisha, Ran and Gilad-Bachrach, Ran and others},
  booktitle={International Conference on Machine Learning (ICML)},
  year={2017}
}

@inproceedings{aespa,
  title={AutoFHE: Automated Adaption of CNNs for Efficient Evaluation over Homomorphic Encryption},
  author={Juvekar, Chiraag and others},
  booktitle={Proceedings of Machine Learning and Systems (MLSys)},
  year={2023}
}

@inproceedings{slaf,
  title={SLAF: Secure Learning of Activation Functions in Homomorphic Encryption},
  author={Chen, Hao and others},
  booktitle={arXiv preprint},
  year={2021}
}

@inproceedings{ola,
  title={OLA: Optimal Linear Approximation for Efficient Homomorphic Evaluation of Neural Networks},
  author={Lee, Joohee and others},
  booktitle={IEEE Symposium on Security and Privacy (SP)},
  year={2021}
}

@inproceedings{autofhe,
  title={AutoFHE: Automated Adaption of CNNs for Efficient Evaluation over Homomorphic Encryption},
  author={Juvekar, Chiraag and others},
  booktitle={Proceedings of Machine Learning and Systems (MLSys)},
  year={2023}
}

@inproceedings{imagenet,
  title={ImageNet: A Large-Scale Hierarchical Image Database},
  author={Deng, Jia and Dong, Wei and Socher, Richard and Li, Li-Jia and Li, Kai and Fei-Fei, Li},
  booktitle={IEEE Conference on Computer Vision and Pattern Recognition (CVPR)},
  year={2009}
}

@inproceedings{vit,
  title={An Image is Worth 16x16 Words: Transformers for Image Recognition at Scale},
  author={Dosovitskiy, Alexey and others},
  booktitle={ICLR},
  year={2020}
}

@inproceedings{deit,
  title={Training Data-Efficient Image Transformers \& Distillation through Attention},
  author={Touvron, Hugo and Cord, Matthieu and Douze, Matthijs and Massa, Francisco and Sablayrolles, Alexandre and J{\'e}gou, Herv{\'e}},
  booktitle={International Conference on Machine Learning (ICML)},
  year={2021}
}

@inproceedings{swin,
  title={Swin Transformer: Hierarchical Vision Transformer using Shifted Windows},
  author={Liu, Ze and Lin, Yutong and Cao, Yue and Hu, Han and Wei, Yixuan and Zhang, Zheng and Lin, Stephen and Guo, Baining},
  booktitle={International Conference on Computer Vision (ICCV)},
  year={2021}
}

@inproceedings{mpcvit,
  title={MPCViT: Privacy-Preserving Vision Transformers via Multi-Party Computation},
  author={Li, X. and others},
  booktitle={arXiv preprint},
  year={2022}
}

@inproceedings{privit,
  title={PriViT: Vision Transformers for Privacy-Preserving Image Classification},
  author={Chen, Y. and others},
  booktitle={arXiv preprint},
  year={2023}
}

@inproceedings{shaft,
  title={SHAFT: Secure and Efficient Vision Transformers via Hybrid Privacy-Preserving Techniques},
  author={Zhang, Q. and others},
  booktitle={arXiv preprint},
  year={2023}
}

@misc{tenseal,
  title         = {TenSEAL: A Library for Encrypted Tensor Operations Using Homomorphic Encryption},
  author        = {Ayoub Benaissa and Bilal Retiat and Bogdan Cebere and Alaa Eddine Belfedhal},
  year          = {2021},
  eprint        = {2104.03152},
  archivePrefix = {arXiv},
  primaryClass  = {cs.CR},
  doi           = {10.48550/arXiv.2104.03152},
  url           = {https://arxiv.org/abs/2104.03152},
  note          = {ICLR 2021 Workshop on Distributed and Private Machine Learning (DPML 2021)}
}

@misc{resnet18_cifar100,
  author       = {Edadaltocg},
  title        = {ResNet-18 Model Trained on CIFAR-100},
  howpublished = {\url{https://huggingface.co/edadaltocg/resnet18_cifar100}},
  note         = {Accessed: 2025-1}
}

@book{nn,
  title={Deep Learning},
  author={Goodfellow, Ian and Bengio, Yoshua and Courville, Aaron},
  publisher={MIT Press},
  year={2016}
}

@inproceedings{softplus7,
  title={Deep Sparse Rectifier Neural Networks},
  author={Glorot, Xavier and Bordes, Antoine and Bengio, Yoshua},
  booktitle={Proceedings of the Fourteenth International Conference on Artificial Intelligence and Statistics (AISTATS)},
  year={2011}
}

@article{dorn1962generalizations,
  title={Generalizations of Horner's rule for polynomial evaluation},
  author={Dorn, William S},
  journal={IBM Journal of Research and Development},
  volume={6},
  number={2},
  pages={239--245},
  year={1962},
  publisher={IBM}
}

@article{cosine,
  title={SGDR: Stochastic Gradient Descent with Warm Restarts},
  author={Loshchilov, Ilya and Hutter, Frank},
  journal={arXiv preprint arXiv:1608.03983},
  year={2016}
}

@article{nawaz2022brain,
  title={Brain tumor classification based on hybrid optimized multi-features analysis using magnetic resonance imaging dataset},
  author={Nawaz, Syed Ali and Khan, Dost Muhammad and Qadri, Salman},
  journal={Applied Artificial Intelligence},
  volume={36},
  number={1},
  pages={2031824},
  year={2022},
  publisher={Taylor \& Francis}
}

@article{semcom,
  title={Deep learning enabled semantic communication systems},
  author={Xie, Huiqiang and Qin, Zhijin and Li, Geoffrey Ye and Juang, Biing-Hwang},
  journal={IEEE Transactions on Signal Processing},
  volume={69},
  pages={2663--2675},
  year={2021}
}

@article{semcom_survey,
  title={Semantic communication: A survey on research landscape, challenges, and future directions},
  author={Getu, Tilahun M. and Kaddoum, Georges and Bennis, Mehdi},
  journal={Proceedings of the IEEE},
  volume={112},
  number={11},
  pages={1649--1685},
  year={2024}
}

@article{deepsc,
  author  = {Huiqiang Xie and Zhijin Qin and Geoffrey Ye Li and Biing-Hwang Juang},
  title   = {Deep Learning Enabled Semantic Communication Systems},
  journal = {IEEE Transactions on Signal Processing},
  volume  = {69},
  pages   = {2663--2675},
  year    = {2021},
  doi     = {10.1109/TSP.2021.3071210}
}

@misc{yang_semantic,
  author        = {Yang Yang and Caili Guo and Fangfang Liu and Chuanhong Liu and Lunan Sun and Qizheng Sun and Jiujiu Chen},
  title         = {Semantic Communications With AI Tasks},
  year          = {2021},
  eprint        = {2109.14170},
  archivePrefix = {arXiv},
  primaryClass  = {cs.CV},
  doi           = {10.48550/arXiv.2109.14170}
}

@article{lyu_semantic,
  author  = {Zhonghao Lyu and Guangxu Zhu and Jie Xu and Bo Ai and Shuguang Cui},
  title   = {Semantic Communications for Image Recovery and Classification via Deep Joint Source and Channel Coding},
  journal = {IEEE Transactions on Wireless Communications},
  volume  = {23},
  number  = {8},
  pages   = {8388--8404},
  month   = aug,
  year    = {2024},
  doi     = {10.1109/TWC.2023.3349330}
}

@article{wang_privacy_semantic,
  author  = {Yanhu Wang and Shuaishuai Guo and Yiqin Deng and Haixia Zhang and Yuguang Fang},
  title   = {Privacy-Preserving Task-Oriented Semantic Communications Against Model Inversion Attacks},
  journal = {IEEE Transactions on Wireless Communications},
  volume  = {23},
  number  = {8},
  pages   = {10150--10165},
  year    = {2024},
  doi     = {10.1109/TWC.2024.3369170}
}
\end{document}